\documentclass[twocolumn,prl,aps,showpacs]{revtex4}
\usepackage{graphicx}

\begin{document}

\newcommand{\snn}{\sqrt{s_{NN}}}
\newcommand{\seff}{\sqrt{s_{\rm eff}}}
\newcommand{\s}{\sqrt{s}}
\newcommand{\pp}{pp}
\newcommand{\pbarp}{\overline{p}p}
\newcommand{\qbarq}{\overline{q}q}
\newcommand{\epem}{e^+e^-}
\newcommand{\nhit}{N_{hit}}
\newcommand{\npp}{n_{pp}}
\newcommand{\nch}{N_{ch}}
\newcommand{\avench}{\langle\nch\rangle}
\newcommand{\np}{N_{part}}
\newcommand{\ns}{N_{spec}}
\newcommand{\ntot}{\langle\nch\rangle}
\newcommand{\avenp}{\langle\np\rangle}
\newcommand{\npB}{N_{part}^B}
\newcommand{\nc}{N_{coll}}
\newcommand{\avenc}{\langle\nc\rangle}
\newcommand{\half}{\frac{1}{2}}
\newcommand{\halfnp}{\langle\np/2\rangle}
\newcommand{\etap}{\eta^{\prime}}
\newcommand{\as}{\alpha_{s}(s)}
\newcommand{\etazero}{\eta = 0}
\newcommand{\etaone}{|\eta| < 1}
\newcommand{\dndeta}{d\nch/d\eta}
\newcommand{\dndetazero}{\dndeta|_{\etazero}}
\newcommand{\dndetaone}{\dndeta|_{\etaone}}
\newcommand{\dndetanp}{\dndeta / \halfnp}
\newcommand{\dndetaonp}{\dndeta / \np}
\newcommand{\dndetazeronp}{\dndetazero / \halfnp}
\newcommand{\dndetaonenp}{\dndetaone / \halfnp}
\newcommand{\ratio}{\ntot/\halfnp}
\newcommand{\nee}{N_{\epem}}
\newcommand{\nhh}{N_{hh}}
\newcommand{\nubar}{\overline{\nu}}
\newcommand{\stc}{\sigma^2_C}
\newcommand{\stcr}{\sigma^2_{C,raw}}

\title{Forward-Backward Multiplicity Correlations in $\snn=200$ GeV Au+Au Collisions}

\author{
%
%
B.B.Back$^1$,
M.D.Baker$^2$,
M.Ballintijn$^4$,
D.S.Barton$^2$,
R.R.Betts$^6$,
A.A.Bickley$^7$,
R.Bindel$^7$,
A.Budzanowski$^3$,
W.Busza$^4$,
A.Carroll$^2$,
Z.Chai$^2$,
M.P.Decowski$^4$,
E.Garc\'{\i}a$^6$,
T.Gburek$^3$,
N.George$^{1,2}$,
K.Gulbrandsen$^4$,
S.Gushue$^2$,
C.Halliwell$^6$,
J.Hamblen$^8$,
M.Hauer$^2$,
G.A.Heintzelman$^2$,
C.Henderson$^4$,
D.J.Hofman$^6$,
R.S.Hollis$^6$,
R.Ho\l y\'{n}ski$^3$,
B.Holzman$^2$,
A.Iordanova$^6$,
E.Johnson$^8$,
J.L.Kane$^4$,
J.Katzy$^{4,6}$,
N.Khan$^8$,
W.Kucewicz$^6$,
P.Kulinich$^4$,
C.M.Kuo$^5$,
W.T.Lin$^5$,
S.Manly$^8$,
D.McLeod$^6$,
A.C.Mignerey$^7$,
R.Nouicer$^6$,
A.Olszewski$^3$,
R.Pak$^2$,
I.C.Park$^8$,
H.Pernegger$^4$,
C.Reed$^4$,
L.P.Remsberg$^2$,
M.Reuter$^6$,
C.Roland$^4$,
G.Roland$^4$,
L.Rosenberg$^4$,
J.Sagerer$^6$,
P.Sarin$^4$,
P.Sawicki$^3$,
H.Seals$^2$,
I.Sedykh$^2$,
W.Skulski$^8$,
C.E.Smith$^6$,
M.A.Stankiewicz$^2$,
P.Steinberg$^2$,
G.S.F.Stephans$^4$,
A.Sukhanov$^2$,
J.-L.Tang$^5$,
M.B.Tonjes$^7$,
A.Trzupek$^3$,
C.Vale$^4$,
G.J.van~Nieuwenhuizen$^4$,
S.S.Vaurynovich$^4$,
R.Verdier$^4$,
G.I.Veres$^4$,
E.Wenger$^4$,
F.L.H.Wolfs$^8$,
B.Wosiek$^3$,
K.Wo\'{z}niak$^3$,
A.H.Wuosmaa$^1$,
B.Wys\l ouch$^4$\\
\small
%
%
%
%
$^1$~Argonne National Laboratory, Argonne, IL 60439-4843, USA\\
$^2$~Brookhaven National Laboratory, Upton, NY 11973-5000, USA\\
$^3$~Institute of Nuclear Physics PAN, Krak\'{o}w, Poland\\
$^4$~Massachusetts Institute of Technology, Cambridge, MA 02139-4307, USA\\
$^5$~National Central University, Chung-Li, Taiwan\\
$^6$~University of Illinois at Chicago, Chicago, IL 60607-7059, USA\\
$^7$~University of Maryland, College Park, MD 20742, USA\\
$^8$~University of Rochester, Rochester, NY 14627, USA\\
}
\date{\today}

\begin{abstract}
Forward-backward correlations of charged-particle multiplicities in symmetric bins
in pseudorapidity are studied in order to gain insight into the
underlying correlation structure of particle production in Au+Au collisions.
The PHOBOS detector is used to measure integrated multiplicities
in bins centered at $\eta$, defined within $|\eta|<3$, and covering
intervals $\Delta \eta$.  The variance $\stc$ of
a suitably defined forward-backward asymmetry variable $C$
is calculated as a function of
$\eta$, $\Delta\eta$, and centrality.  It is found to be sensitive to short
range correlations, and the concept of ``clustering'' is used to
interpret comparisons to phenomenological models.

\end{abstract}
\maketitle

At the top energies reached for Au+Au collisions at 
the Relativistic Heavy Ion 
Collider ($\snn=200$ GeV), most of the energy is carried by particles
with large longitudinal momenta.  In general, these momenta are best
expressed with the rapidity variable ($y=\frac{1}{2}\ln \frac{E+p_z}{E-p_z}$)
or its near equivalent, the pseudorapidity variable 
($\eta = -\ln(\tan(\theta/2))$).
In strongly interacting systems, it is often
thought that correlations between two particles
are mainly ``short-range'' in rapidity.
Indeed, short range correlations have been observed in $pp$ and $\pbarp$ collisions
over a wide range of beam energies \cite{Ansorge:1988fg} 
via two-particle correlation measurements.
However, they have never been shown to be the only source of correlation
in multiparticle production.

Indeed, single-particle rapidity (or pseudorapidity)
distributions for inclusive charged particles produced
in heavy ion collisions reveal a distinct ``trapezoidal'' 
structure stretching across
the full rapidity range available in these reactions.  
More importantly,
there is no evidence for an extended boost invariance, which would 
imply independent emission from different rapidity regions.  Rather,
two non-trivial effects are visible in the centrality
dependence of particle production, which are apparently long-range
in rapidity.  The first is the fact that integrating over the full phase
space reveals that particle production depends linearly on the number of
participants~\cite{Back:2003xk}.  
The second is that this occurs despite a significant change in the shape
of the pseudorapidity dependence as a function of centrality, 
which happens to be collision-energy independent in the forward 
region~\cite{Back:2002wb}.  

All of this suggests that charged-particle 
production is highly correlated
over large regions of rapidity, which
naturally begs the question of the underlying structure of the 
single-particle distributions.  
In this paper, we discuss how to 
take first steps in this direction via the study of forward-backward
multiplicity correlations.
By this is meant the event-by-event
comparison of the integrated multiplicity  $N_F$ in a bin defined in the
forward ($\eta>0$) region, centered at $\eta$ with pseudorapidity 
interval $\Delta\eta$,
with the multiplicity $N_B$ measured in an identical bin defined 
in the backward hemisphere, centered at $-\eta$.  
With these definitions, one
can construct the event-wise observable $C=(N_F-N_B)/\sqrt{N_F+N_B}$,
and measure the variance $\sigma^2_C$ for a set of events with nominally
similar characteristics (e.g. collision centrality).

The $C$ variable is chosen to have particular sensitivity to various types
of long and short range correlations.  An ``intrinsic'' long-range
correlation~\cite{Alpgard:1983xp} in the emission of particles
into the forward {\it and} backward hemisphere from a single
source would give $N_F-N_B = 0$ with a substantial
value of $N_F+N_B$, forcing $C$ to $0$.  
However, if the particle sources tended to produce into the forward
{\it or} backward region, such that the partitioning was binomial, this
would lead to $\stc=1$, since $\sigma^2(N_F-N_B)=N_F+N_B$ in that case.
Direct studies of $\langle N_F \rangle$ as a function of 
$N_B$ were used by UA5 for $\pbarp$ collisions to conclude
that there are no large intrinsic correlations in particle production.
And yet,
a non-trivial long range correlation between the hemispheres persists,
even when excluding the central two units of 
pseudorapidity~\cite{Ansorge:1988fg,Alpgard:1983xp}.

Short-range correlations would arise if objects emitted into
either hemisphere break into $\langle k \rangle$ particles on average, 
with a variance of $\sigma^2_k$, each of which
remains close in rapidity (e.g. due to isotropic emission).  Such
intrinsic short-range correlations have in fact been measured
in $\pbarp$ and $pp$ experiments by direct construction of the two-particle
correlation function in $\eta$.  In a series of papers, the
UA5 collaboration explored the hypothesis that particle emission
was dominated by the decay of ``clusters'' locally
in rapidity space~\cite{Ansorge:1988fg,Alpgard:1983xp}.  
By varying $\Delta\eta$ and measuring the ratio
$4\sigma^2_F/\langle N_F \rangle$ as a function of $N_F+N_B$,
they found an ``effective'' cluster multiplicity ($k_{eff}=\langle k \rangle
+ \sigma^2_k / \langle k \rangle$) of approximately 2 charged
particles.  
This exceeds the multiplicity expected from a resonance
gas using masses up to 1.5 GeV, which gives an average multiplicity
per particle of about $\langle k \rangle \sim 1.5$~\cite{Stephanov:1999zu}, 
suggesting that
clusters may have a variety of dynamical sources (although, to be sure
of this, resonances over a wider mass range 
than in the previous calculations should be taken into account).

The decay of clusters has a particular effect on $\stc$.  
Consider the idealized case,
where a cluster decays into exactly $k$ particles,
and all of the $N$ particles are randomly distributed into
identical $\eta$-bins in the forward or backward directions.
In this case, the underlying
fluctuations are those associated with $N/k$ objects rather than
$N$ independent particles, and $C\rightarrow \sqrt{k} C$.
In other words, $\stc$ is linear with the cluster multiplicity
for this simple case.
In a more realistic physical situation, particles from each cluster may
fall outside the chosen $\eta$-bin, or even in the opposite
hemisphere.  This results in a non-trivial modification of 
the measured value of $k$, but one which is still related directly to
$k$, or rather $k_{eff}$ (which incorporates the effect of the multiplicity
per cluster having its own distribution).
This relationship can be explored with the use of simulations
and thus used to extract the effective cluster multiplicity with
measured forward-backward correlations.
These studies show that
forward-backward correlations have the surprising ability to
measure a fundamental {\it local} property of particle production. 

\begin{figure}[t]
\begin{center}
\includegraphics[width=80mm]{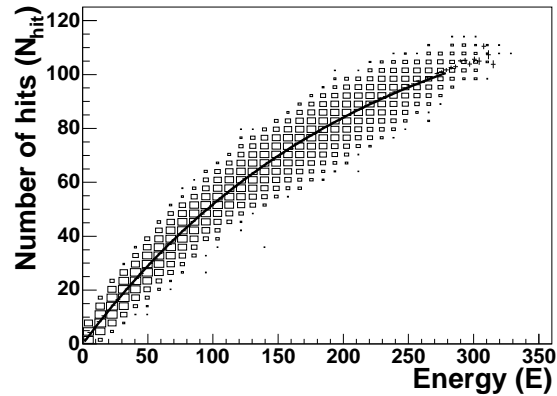}
\end{center}
\caption{Number of hits vs. total deposited energy (in keV) measured $-0.5<\eta<0$ (in the Octagon detector) for 200 GeV Au+Au collisions.   The solid line is a function, described in the text, fit to the mean number of hits in each energy bin.
\label{fig:NE}}
\end{figure}

The data analyzed here were taken with the PHOBOS 
detector~\cite{Back:2003sr} during 
Runs 2 and 4, in 2001 and 2004, respectively.  The pseudorapidity 
acceptance was restricted to that of the ``Octagon'' detector,
which is a tube of silicon sensors covering $|\eta|<3$ and
full azimuth except for a region near midrapidity.  To simplify
the analysis for different values of $\eta$ and $\Delta \eta$, 
only the regions of the detector with complete rapidity coverage were kept, 
restricting the total azimuthal acceptance to $\Delta \phi = \pi$,
in four 45-degree wedges~\cite{Wozniak:2004kp}.
The multiplicity in each bin is estimated event-by-event 
by summing up the angle-corrected deposited energy of all
hits and then dividing by the average energy per particle~\cite{Chai:2005}.
The average is calculated by a fit to the 
distribution of $\langle N_{hits} \rangle$
vs. the total deposited energy (E), an example of which is shown
in Fig.~\ref{fig:NE}, with the function
$\langle N_{hit} \rangle = N_{max} (1-e^{-E/E_{max}})$.  The average 
energy per particle is given by $E_{max}/N_{max}$.
This removes nearly all occupancy-related effects.
Beyond the usual lower threshold to define a hit in the silicon,
an $\eta$-dependent upper bound of the deposited energy per hit
is applied to reduce the effect of low-momentum 
secondaries on the fluctuations
of the single-particle $dE/dx$ distributions, while keeping
$97\%$ of the primaries.
The average $\langle C \rangle$, calculated as a function of
$\eta$, centrality, and event vertex, is subtracted event-by-event
to correct for gaps in the Octagon.
This correction was studied in detail with simulations, and 
was found to leave the fluctuations unaffected.

\begin{figure}[t]
\begin{center}
\includegraphics[width=80mm]{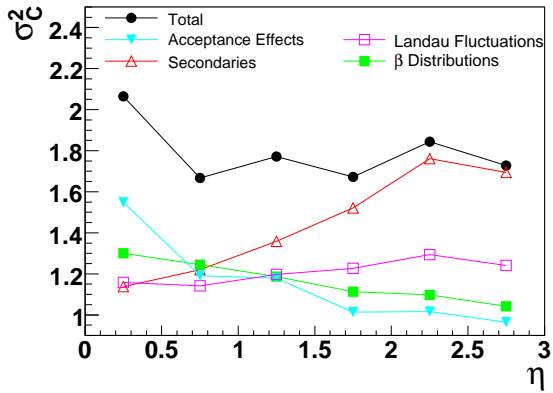}
\end{center}
\caption{Contributions to $\stc$ from sources other than primary
charged particles.
\label{fig:sources}}
\end{figure}

To provide information that can be compared directly to models, a
procedure was developed to estimate and remove the detector 
effects from the raw measured value of $\stc$ ($\stcr$) 
by using Monte Carlo (MC) 
simulations of the PHOBOS apparatus.  The basic idea is to
assume that $\stcr = \stc + \sigma^2_{det}$, where
$\sigma^2_{det}$ is the contribution from detector effects, 
and use the MC to subtract it on average, leaving
behind only the correlations between primary particles.  
It is found that
there are several sources which contribute differently as a
function of $\eta$ and combine in quadrature to a nearly-constant value
over the pseudorapidity range covered by the Octagon~\cite{Chai:2005},
as illustrated in Fig.~\ref{fig:sources}.  Gaps between the silicon sensors
tend to be most important near mid-rapidity and decrease rapidly with
angle as the particle density decreases.  Similarly, larger-angle particles
tend to be slower than those at forward angles, leading to larger 
fluctuations in the energy deposition in the single layer of silicon.
At the same time, the
number of secondary particles emitted from primaries interacting in the 
beryllium beampipe rises quickly with increasing rapidity.  This
leads to increased fluctuations, even though some of the background
is removed with a properly-chosen lower cut on
the angle-corrected $dE/dx$ of each hit.
We have also studied fluctuations from the momentum (or $\beta$)
distributions of the particles incident on the detector.  In this case,
the contribution to $\stc$  decreases with increasing rapidity.

\begin{figure}[t]
\begin{center}
\includegraphics[width=80mm]{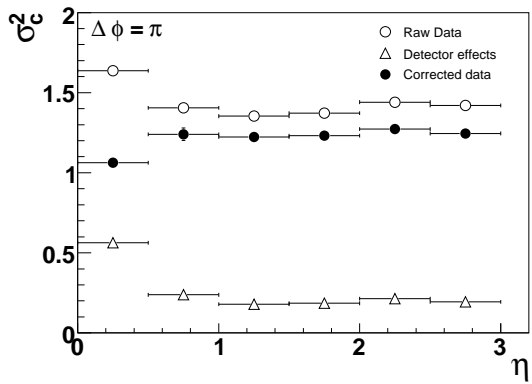}
\end{center}
\caption{Example of the analysis method, showing the raw
values of $\stc$ ($\stcr$), the contribution from
detector effects ($\sigma^2_{det}$, as estimated using
modified HIJING), and the subtracted
result ($\stc$) for bins of $\Delta \eta=0.5$
as a function of $\eta$ in the Run 4 data set.
Note that these values are calculated for the half-azimuth
($\Delta\phi = \pi$) acceptance, which is corrected-for in
the final result.
\label{fig:s2c_corr_pas}}
\end{figure}

The detector effects are corrected on average by calculating $\stc$ using a
modified HIJING simulation with all intrinsic correlations
destroyed by flipping the sign of $\eta$ at random, effectively breaking
any rapidity-dependent multi-particle correlations and forcing $\stc = 1$.
This allows a direct estimation of $\sigma^2_{det}$.
An example of this is shown in Fig.~\ref{fig:s2c_corr_pas},
which shows the raw
values of $\stc$ ($\stcr$), the contribution from
detector effects ($\sigma^2_{det}$, as estimated from
modified HIJING), and the subtracted
result ($\stc$) for bins of $\Delta \eta=0.5$
as a function of $\eta$ in the Run 4 data set.
There is a small residual correlation of $\sigma^2_{det}$ with
$\stc$, parametrized as $\sigma^2_{det}=\sigma^{2}_{det,raw}(1-\alpha 
(\stc -1 ))$, which is removed by estimation of $\alpha$
as a function of $\eta$, $\Delta\eta$ and centrality.
Finally, we correct for using
only half-azimuth acceptance by $\stc \rightarrow 2(\stc-1)+1$, 
a formula obtained using MC simulations, 
assuming that the
limited acceptance only reduces the probability of measuring
all of the particles in a typical cluster.

\begin{figure}[t]
\begin{center}
\includegraphics[width=80mm]{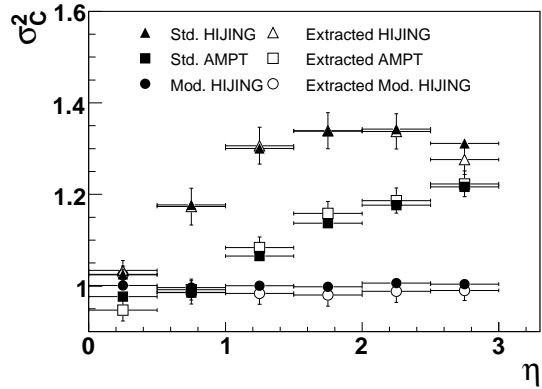}
\end{center}
\caption{
Calibration of reconstruction method with MC simulations, as
a function of $\eta$ with $\Delta \eta=0.5$.  Solid points represent
MC truth information on $\stc$, while the open symbols represent the outcome
of the reconstruction procedure.
\label{fig:ptest1}}
\end{figure}

Systematic uncertainties were calculated by varying several variables 
involved the estimation of $\stc$ (data sets, event generators, and $dE/dx$ cuts)
and found to be around
$\Delta \stc \sim 0.1$.  The bin-to-bin variation of the systematic
error calculation was adjusted to reduce fluctuations
from the error determination procedure.

After correcting for detector effects, the results on $\stc$ can be
directly compared with model calculations based on charged primary particles.
We have focused mainly on HIJING~\cite{Gyulassy:1994ew} 
and AMPT~\cite{Lin:2004en} (which includes partonic and hadronic transport
codes), both of which have been used to
describe various features of heavy ion collisions at RHIC energies.
Fig.~\ref{fig:ptest1} shows the results of the correction procedure
described above on three simulations.  Two of them are based on standard
versions of HIJING and AMPT, and a third on the modified HIJING, as
described previously.
One sees that in all three cases, the reconstructed values of $\stc$, using
the same tuning in all cases,
match the $\stc$ extracted from the primary charged particles
within statistical precision.

\begin{figure*}[t]
\begin{center}
\includegraphics[width=140mm]{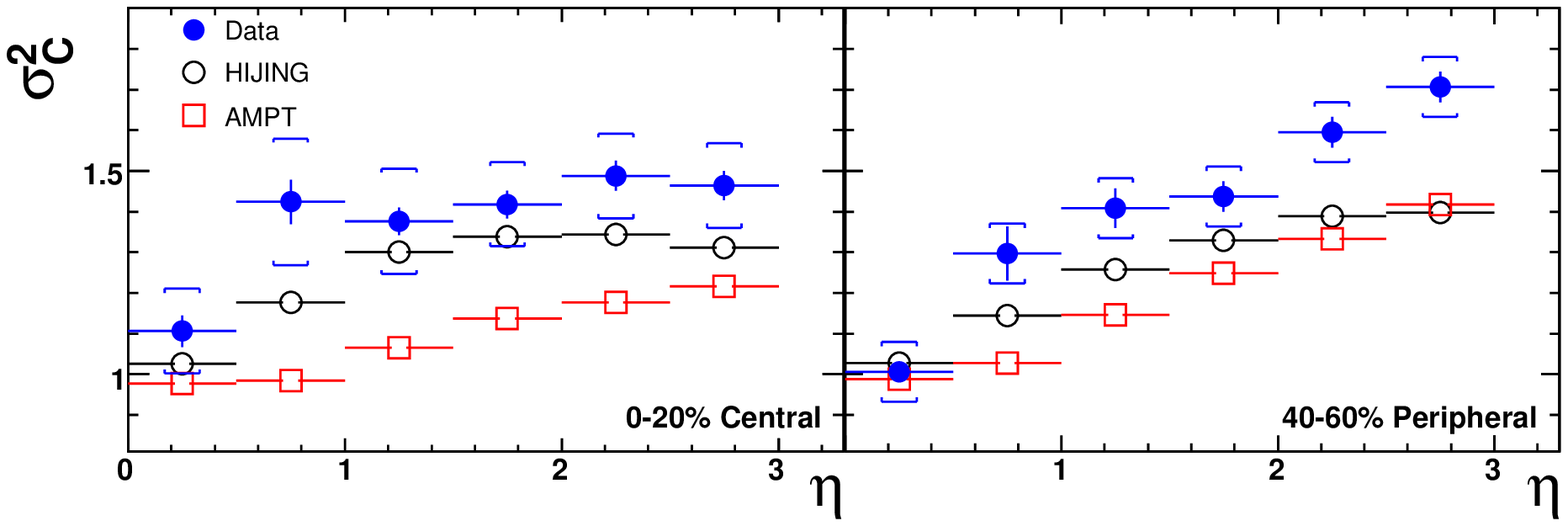}
\includegraphics[width=140mm]{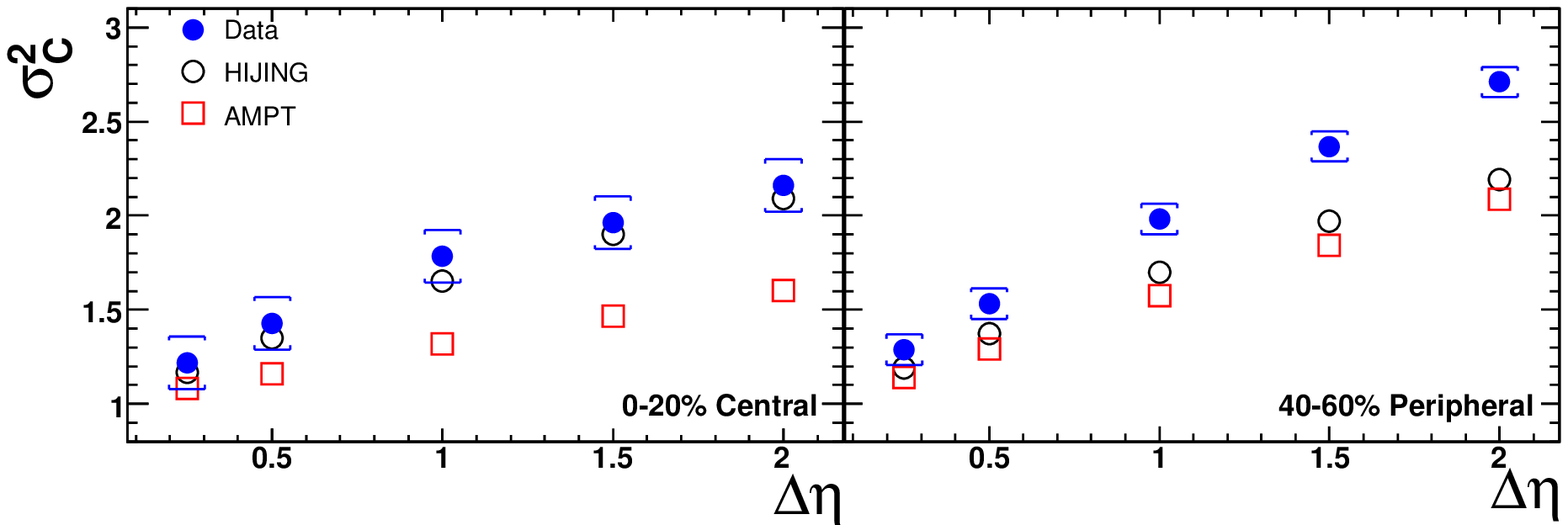}
\end{center}
\caption{(upper) $\stc$ for a fixed $\Delta \eta=0.5$ as a function of $\eta$.  (lower) $\stc$ as a function of $\Delta \eta$ for a fixed
bin center at $\eta=2.0$.  Statistical errors are shown as vertical bars, 
while systematic errors are shown as brackets.
\label{fig:etadeta}}
\end{figure*}

The first set of results concerns the $\eta$ dependence of $\stc$, for
forward and backward bins 
that are $\Delta \eta=0.5$ units wide, as shown in the upper panel
of Fig.~\ref{fig:etadeta}.
In this case, HIJING and AMPT already show some 
differences related to their underlying physics scenarios.  
For peripheral (40-60\%) events, both models 
have a similar magnitude and a monotonically-rising $\eta$ dependence.
Central events show a substantial difference extending over most of
the rapidity range, with AMPT showing a systematically smaller 
value of $\stc$.  This may be due to the initially produced clusters
being smeared out in rapidity space 
by the hadronic rescattering stage.  Also, it is
observed that both data and MC show $\stc \sim 1$ at $\eta = 0$.  
This is suggestive of the effect predicted by Jeon and Shi
\cite{Shi:2005rc},
that the formation of a QGP near mid-rapidity
should break up any sort of cluster structure seen in $\pp$ and thus
lead to a reduction in $\stc$.
However,
it can also be explained by the fact that clusters produced at $\eta=0$ tend 
to emit particles into
{\it both} the $N_F$ and $N_B$ side, inducing an ``intrinsic'' long-range
correlation that decreases $\stc$.  Simulations show that this effect is not
modified by the precise values of $\langle k \rangle$ and $\sigma^2_k$,
for constant $k_{eff}$.
In any case, the data are systematically higher than
the model calculations for both peripheral and central events.

The next set of results is the $\Delta \eta$ dependence of $\stc$,
with a fixed bin position $\eta = 2.0$, as shown in the lower panel of
Fig. \ref{fig:etadeta}.  
One sees that for both peripheral and
central data, $\stc$ rises monotonically with increasing pseudorapidity
interval.  This can be explained as an acceptance effect in the
context of cluster emission: by increasing $\Delta \eta$, one
increases the probability of observing more than one particle emitted
from a single cluster in either $N_F$ or $N_B$.  Clearly, the rate of change
with $\Delta \eta$ should reflect the full cluster distribution
(both $\langle k \rangle$ and $\sigma_k$), but these studies are
not yet sufficiently precise to determine the detailed properties of
clusters produced in RHIC collisions.
It is striking that the peripheral
data has already reached $\stc \sim 3$ for the largest $\Delta \eta$
while this number is closer to 2 for central
data.  Finally, it is interesting that neither HIJING nor AMPT can
explain both the centrality and $\Delta\eta$ dependence simultaneously.
AMPT never agrees with the data in magnitude, but at least
predicts a larger $\stc$ in peripheral events than in central
events.  This should be contrasted with HIJING, which
reproduces the central data but has no centrality dependence at all,
at odds with the experimental data.

In conclusion, measurements of forward-backward fluctuations 
of charged-particles produced in Au+Au collisions at $\snn=200$ GeV
may provide insight into the structure of long and short range 
correlations in pseudorapidity space.  Our data for
200 GeV Au+Au collisions are now fully corrected for detector 
and background effects, so direct comparisons can be made to
phenomenological models.  We see significant short-range
correlations at all centralities and pseudorapidities, instead of
just at mid-rapidity.  There is a non-trivial centrality and
rapidity dependence of these correlations, in both $\eta$ and
$\Delta\eta$.  Finally, neither HIJING nor AMPT reproduces all of
the main qualitative features, but the way in which they fail to do so
may well provide information on the underlying physics.  In
particular, more theoretical attention should be paid to the
properties of ``clusters'' required to explain our data.  
As mentioned above, Jeon and Shi have proposed
that QGP formation would modify the measured properties of 
clusters~\cite{Shi:2005rc}.
The data shown here should provide means to 
study such effects, or set upper limits on their occurrence.

This work was partially supported by U.S. DOE grants 
DE-AC02-98CH10886,
DE-FG02-93ER40802, 
DE-FC02-94ER40818,  
DE-FG02-94ER40865, 
DE-FG02-99ER41099, and
W-31-109-ENG-38, by U.S. 
NSF grants 9603486, 
0072204,            
and 0245011,        
by Polish KBN grant 1-P03B-062-27(2004-2007),
by NSC of Taiwan Contract NSC 89-2112-M-008-024, and
by Hungarian OTKA grant (F 049823).

\end{document}